\documentclass[final,authoryear, 5p, times]{elsarticle}

\makeatletter
\def\ps@pprintTitle{%
  \let\@oddhead\@empty
  \let\@evenhead\@empty
  \def\@oddfoot{\reset@font\hfil\thepage\hfil}
  \let\@evenfoot\@oddfoot
}
\makeatother

\usepackage{graphics}
\usepackage{graphicx}

\usepackage{amsmath,amsfonts,amssymb,amsthm}
\usepackage{xcolor}
\usepackage{soul}

\usepackage{newtxtext}
\usepackage[varvw]{newtxmath}

\begin{document}

\begin{frontmatter}



\title{Reduced Hodgkin–Huxley models based on the correlation between sodium and potassium gating variables}

\author{L\'{\i}zia Branco}
\ead{lizia.branco@kit.edu}

\author{Rui Dil\~ao}
\ead{ruidilao@tecnico.ulisboa.pt}
\address{University of Lisbon, Instituto Superior T\'ecnico,
Av. Rovisco Pais, 1049-001 Lisbon, Portugal}

\begin{abstract}
We analyse the interaction between the sodium (Na$^+$) and potassium (K$^+$) gating variables in the 4D Hodgkin-Huxley axonal electrophysiological model. We develop simplified 3D and 2D models that capture the core dynamical behaviour of the original. These models rely on the empirical relationship $h \simeq c(I) - n$ between the gating variables $h$ and $n$, with $c(I)$ depending on stimulus intensity and reflecting an effective correlation between Na$^+$ and K$^+$ gating. Additionally, we derive the relevant cable equations for these reduced models and demonstrate that the action-potential propagation speed is well approximated by $v(R, C_m) = a / (C_m R^{b})$, where $a > 0$, $0 < b < 1$, and the parameters show weak dependence on stimulus strength within the studied range; here, $C_m$ is membrane capacitance, and $R$ is the axon’s internal resistivity. 
\end{abstract}

\begin{keyword}  Sodium and potassium voltage-gated channels    \sep Speed at which action potentials propagate  \sep Solitary action-potential spikes  \end{keyword}


\end{frontmatter}

\section{Introduction}\label{sec:intro}

In 1952, Hodgkin and Huxley presented a mathematical model describing the neuronal response to externally applied electric stimuli under voltage-clamp conditions. The Hodgkin-Huxley (HH) model has significantly advanced our understanding of the mechanisms underlying the generation of neuronal action potentials and their propagation along the axon. The original 4D HH model comprises differential equations that describe the time-dependent behaviour of three types of membrane currents: voltage-gated sodium, voltage-gated potassium, and leak currents. 

Despite its high accuracy in predicting axon dynamics, the four coupled nonlinear equations and numerous parameters of the 4D HH model can hinder an intuitive understanding of the system dynamics and limit its efficiency in modelling neuronal networks. 
To address these challenges, we examined the roles of the individual ionic currents in the dynamics of the 4D HH model and developed new reduced electrophysiological models that describe the neuronal response to external electric stimuli while preserving the model's dynamical properties.
This approach allowed us to investigate the relationship between the action potentials' spiking frequency, the intensity of the electric stimulus, and various cell-specific parameters. Additionally, we explored how axon-internal resistivity influences the speed and shape of action potential spikes.

This paper is organised as follows. In the next section, we derive and compare two reduced 3D and 2D HH  models based on a seemingly interdependent relationship between the sodium and potassium gating variables. Although FitzHugh first reported a relationship between these channels' gates of the form $h \simeq c - n$, where $c$ is a constant (\cite{FH}), the electrophysiological implications were seldom explored in detail, a gap we have addressed in our study.

In section~\ref{sec3}, we introduce the spatial dependence of the reduced 3D and 2D HH models. We assume that the axon comprises a series of channel segments connected by junctions characterised by an internal resistivity parameter (\cite{End}). Subsequently, we compare the dynamic aspects of action potential propagation along the axon of the original 4D HH  model with active Na$^+$ and K$^+$ channels to those of the derived reduced 3D and 2D dynamic models. Additionally, we derive the propagation properties of action potentials as they relate to the axon parameters in response to currents originating from the soma and other external stimuli. The solitary characteristics of certain action potential responses are discussed. Furthermore, we analyse the direction of action potential propagation along the axon when current clamp signals are introduced away from the soma. This provides a simplified representation of localised external stimulation, patch-clamp injection, or activity entering through a branch. In the final section~\ref{sec4}, we summarise and discuss the main conclusions of the paper.

\section{The reduced Hodgkin-Huxley models}\label{sec2}
 
The classical Hodgkin and Huxley model simulates the membrane current of the squid {\it Loligo} as the sum of sodium, potassium, and leak-current components, with independent gating variables for the sodium and potassium channels.
By changing the chemical composition of the extracellular space and evaluating the response of each channel's ionic current, Hodgkin and Huxley developed an empirical model where the specific values of the different parameters were chosen to fit the data best. The 4D HH model  consists of the following four nonlinear ordinary differential equations (\cite{HH1,HH2})
\begin{equation}
\begin{array}{rcl} \displaystyle
   C_m \frac{\mathrm{d} V}{\mathrm{~d} t} &=&I-g_{\hbox{\tiny K}} n^4\left(V-V_{\hbox{\tiny K}}\right)\\ \displaystyle
	&&\displaystyle -g_{\hbox{\tiny Na}} m^3 h\left(V-V_{\hbox{\tiny Na}}\right)-g_{\hbox{\tiny L}}\left(V-V_{\hbox{\tiny L}}\right) \\ \displaystyle
	 \frac{\mathrm{d} n}{\mathrm{d} t} &=&\displaystyle \alpha_n(V)(1-n) - \beta_n(V)n\\ [8pt]\displaystyle
	 \frac{\mathrm{d} m}{\mathrm{d} t} &=& \displaystyle \alpha_m(V)(1-m) - \beta_m(V)m \\ [8pt]\displaystyle
        \frac{\mathrm{d} h}{\mathrm{d} t} &=&\displaystyle \alpha_h(V)(1-h) - \beta_h(V)h
\end{array}
 \label{eq-HH}
    \end{equation}
where
\begin{equation}
\begin{array}{lclrcl}\displaystyle
\alpha_n &=&\displaystyle 0.01 \phi \frac{-V+10}{e^{(-V+10)/10}-1},  &\displaystyle  \beta_n &=&\displaystyle 0.125 \phi e^{-V/80},\\ \displaystyle
\alpha_m &=&\displaystyle 0.1 \phi \frac{-V+25}{e^{(-V+25)/10}-1},  &\displaystyle  \beta_m &=&\displaystyle 4 \phi e^{-V/18},\\ \displaystyle
\alpha_h &=&\displaystyle 0.07 \phi e^{-V/20},  &\displaystyle   \beta_h &=&\displaystyle  \phi \frac{1}{e^{(-V+30)/10}+1},\\ \displaystyle
\phi &=&\displaystyle 3^{(T-6.3)/10}. &&&
\end{array}
\label{eq-HH2}
\end{equation}
In this model, $V=V_{in}-V_{out}$ represents the transmembrane potential difference relative to the potential outside the cells, measured in mV. The variable $I$ is an electric current density in $\mu$A/cm$^2$, and time is measured in milliseconds (ms).  The calibrated Nernst potentials for the squid {\it Loligo} at a temperature $T=6.3^{\circ}$C are  $V_{\hbox{\tiny Na}} = 115$~mV, $V_{\hbox{\tiny K}} = -12$~mV and $V_{\hbox{\tiny L}} = 10.613$~mV. The membrane capacitance is $C_m = 1$~$\mu F/$cm$^2$. The gating variables $n$, $m$, and $h$ describe the activation of the  K$^+$ channel and the activation and inactivation, respectively, of the Na$^+$ channel. 
The parameters $g_{\hbox{\tiny K}} = 36$~mS/cm$^2$, $g_{\hbox{\tiny Na}}=120$~mS/cm$^2$ and $g_{\hbox{\tiny L}} = 0.3$~mS/cm$^2$, where S=$\Omega^{-1}$,  are the maximal potassium, sodium and leak conductances, respectively, and their values were obtained experimentally. 
From this point onward, unless otherwise stated, we set $g_{\hbox{\tiny L}} =0$ in order to focus on the active Na$^+$ and K$^+$ currents.
The current density $I$ describes an external input from the soma to the axon's first segment or a current source applied locally at an axonal segment.  The current density $I$ in the HH model is independent of $V$. The equations describing the propagation along the axon are derived from \eqref{eq-HH}-\eqref{eq-HH2} by adding segments of equal lengths (see section~\ref{sec3}).

The HH equations \eqref{eq-HH}-\eqref{eq-HH2} are derived under the assumption that sodium and potassium gating variables are considered independent.

\begin{figure} 
\centering
\includegraphics[width=0.45\textwidth]{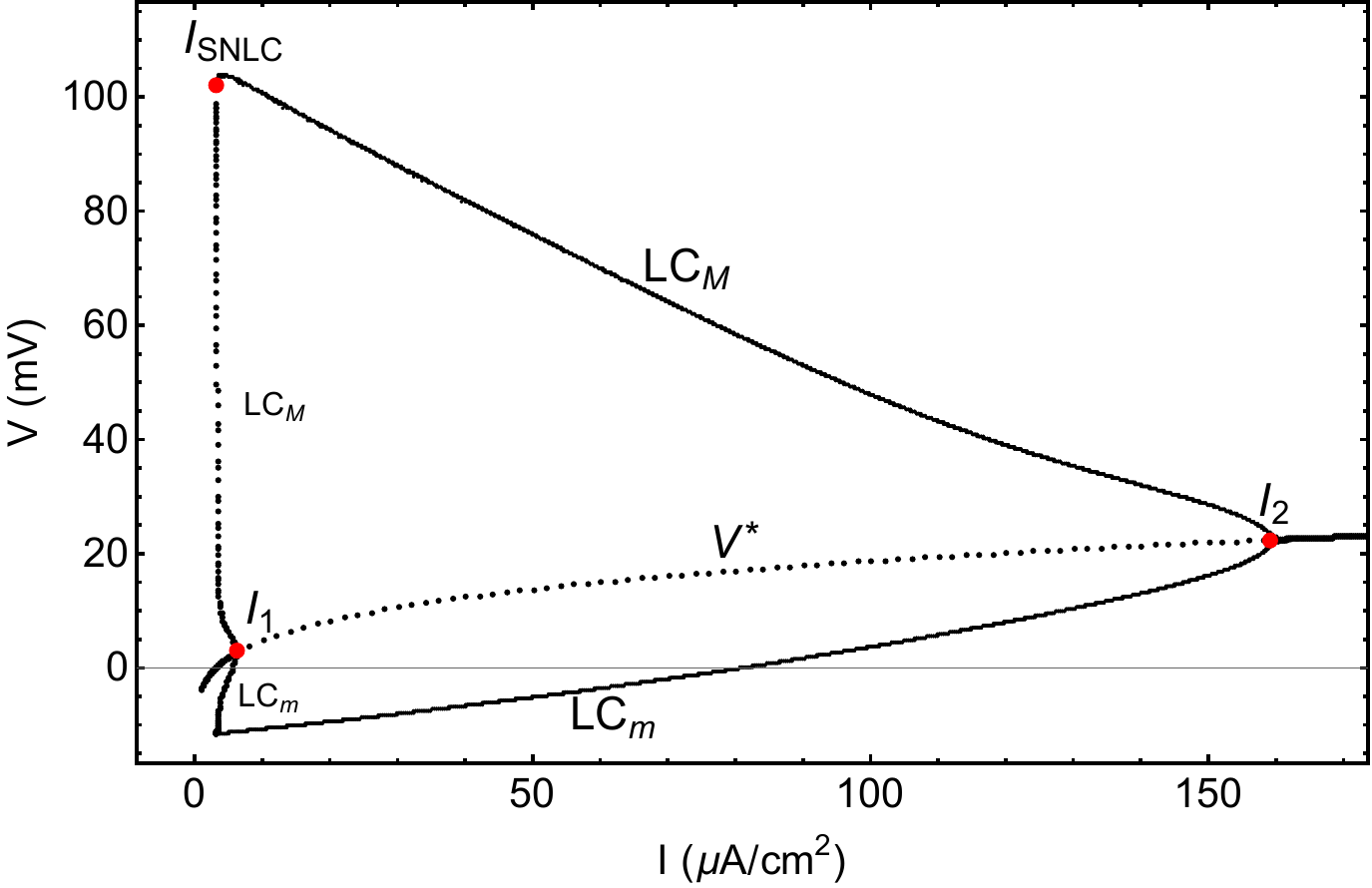}
 \caption{Bifurcation diagram of the 4D HH equations \eqref{eq-HH}-\eqref{eq-HH2} as a function of the electric stimulus $I$ is presented. $LC_M$ and $LC_m$ are the maximum and minimum voltages on a limit cycle.  Parameter values: $C_m = 1$~$\mu F/$cm$^2$,  $V_{\hbox{\tiny Na}} = 115$~mV, $V_{\hbox{\tiny K}} = -12$~mV, $g_{\hbox{\tiny K}} = 36$~mS/cm$^2$, $g_{\hbox{\tiny Na}}=120$~mS/cm$^2$, $g_{\hbox{\tiny L}} = 0.0$~mS/cm$^2$, $I_1 \approx 6.18$~$\mu$A/$\text{cm}^2$, $I_2 \approx 159.20$~$\mu$A/$\text{cm}^2$ and $I_{\hbox{\tiny SNLC}} \approx 3.15$~$\mu$A/$\text{cm} ^2$. Dotted lines correspond to unstable fixed points of the HH model equations, while solid lines represent stable fixed points. Action potential solutions of the 4D HH equations are obtained for $I_{\hbox{\tiny th}}\le I\le I_1$, where $I_{\hbox{\tiny th}}\simeq 2.25$~$\mu$A/$\text{cm}^2$. For $I_1< I<  I_2$, we observe spiking solutions.}
 \label{fig:fig1}
\end{figure}

For each fixed value of $I$, the system of equations \eqref{eq-HH}-\eqref{eq-HH2} has one fixed point with coordinates $p=(V^{*}, n^{*}, m^{*},h^{*})$ that can be stable or unstable. Using the bifurcation analysis software XPPAUT (\cite{Erm}), in Figure~\ref{fig:fig1}, we show the steady state $V^{*}$ as a function of the constant electric stimulus $I$. The two points at which the fixed point changes stability correspond to Hopf bifurcations. The first is subcritical at $I_1 \approx 6.18$~$\mu$A/$\text{cm}^2$, while the second occurs supercritically at $I_2 \approx 159.20$~$\mu$A/$\text{cm}^2$. A limit cycle (LC) is formed at these bifurcation points. In Figure~\ref{fig:fig1}, the dotted LC line corresponds to an unstable limit cycle, while the solid LC line corresponds to a stable limit cycle. The stable and unstable limit-cycle branches coalesce in a saddle-node bifurcation of limit cycles (SNLC), occurring at $I_{\hbox{\tiny SNLC}} \approx 3.15$~$\mu$A/$\text{cm}^2$. The maximum and minimum values of the membrane potential along the limit cycles are labelled as $LC_M$ and $LC_m$, respectively.

Along the first region where $p$ is stable, $I<I_{1}$, for $0\le I<I_{\hbox{\tiny th}}\simeq 2.25$~$\mu$A/$\text{cm}^2$, where $I_{\hbox{\tiny th}}$ is a threshold parameter, all solutions of the 4D HH model converge to the stable steady state $p$. For $I_{\hbox{\tiny th}}<I<I_1$, the solution may exhibit intermittent firing with a finite number of action potential spikes. This behaviour is due to a type I intermittency phenomenon occurring near the SNLC bifurcation, and the number of action potential spikes, $M$, follows the relation $\ln M=C-2\ln(I_{\hbox{\tiny SNLC}} - I)$, where $C$ is a constant (\cite{CD1}). For $I_1 < I < I_2$, the  4D HH model exhibits persistent spiking in response to the stimulus intensities $I$. In the regions $[I_2-\varepsilon_1 , I_2]$, with $\varepsilon_1\simeq 10^{-4}$, the  HH model exhibits  canard-type solutions (\cite{Bra}).  For $I>I_{2}$, $p$ is a stable fixed point.
	
As noted by different authors, the profile of the solution $h(t)$ of equations \eqref{eq-HH}-\eqref{eq-HH2} approximately mirrors $n(t)$, \cite[p. 456]{FH}, \cite{Rin2}, \cite[p. 213]{KS} and \cite{Wan}. 
This property of the solutions suggests the existence of a correlation between the conductances of the Na$^+$ and the K$^+$ channels that Hodgkin and Huxley did not originally consider. Replacing $h$ by $c-n$ in equations \eqref{eq-HH}-\eqref{eq-HH2}, where $c$ is some constant, we obtain the reduced 3D HH model 
\begin{equation}
\begin{array}{rcl}
C_m\frac{dV}{dt} &=& I - g_{\hbox{\tiny K}} n^4(V-V_{\hbox{\tiny K}}) - g_{\hbox{\tiny Na}}m^3(c-n)(V-V_{\hbox{\tiny Na}})\\[8pt] \displaystyle
\frac{dn}{dt}  &=& \displaystyle  \alpha_n(V)(1-n)-\beta_n(V)n\\[8pt]  \displaystyle
\frac{dm}{dt}  &=& \displaystyle  \alpha_m(V)(1-m)-\beta_m(V)m.
\end{array}\label{eq:3dmodel}
\end{equation}

\begin{figure}[!htb]
\centering
\includegraphics[width=0.45\textwidth]{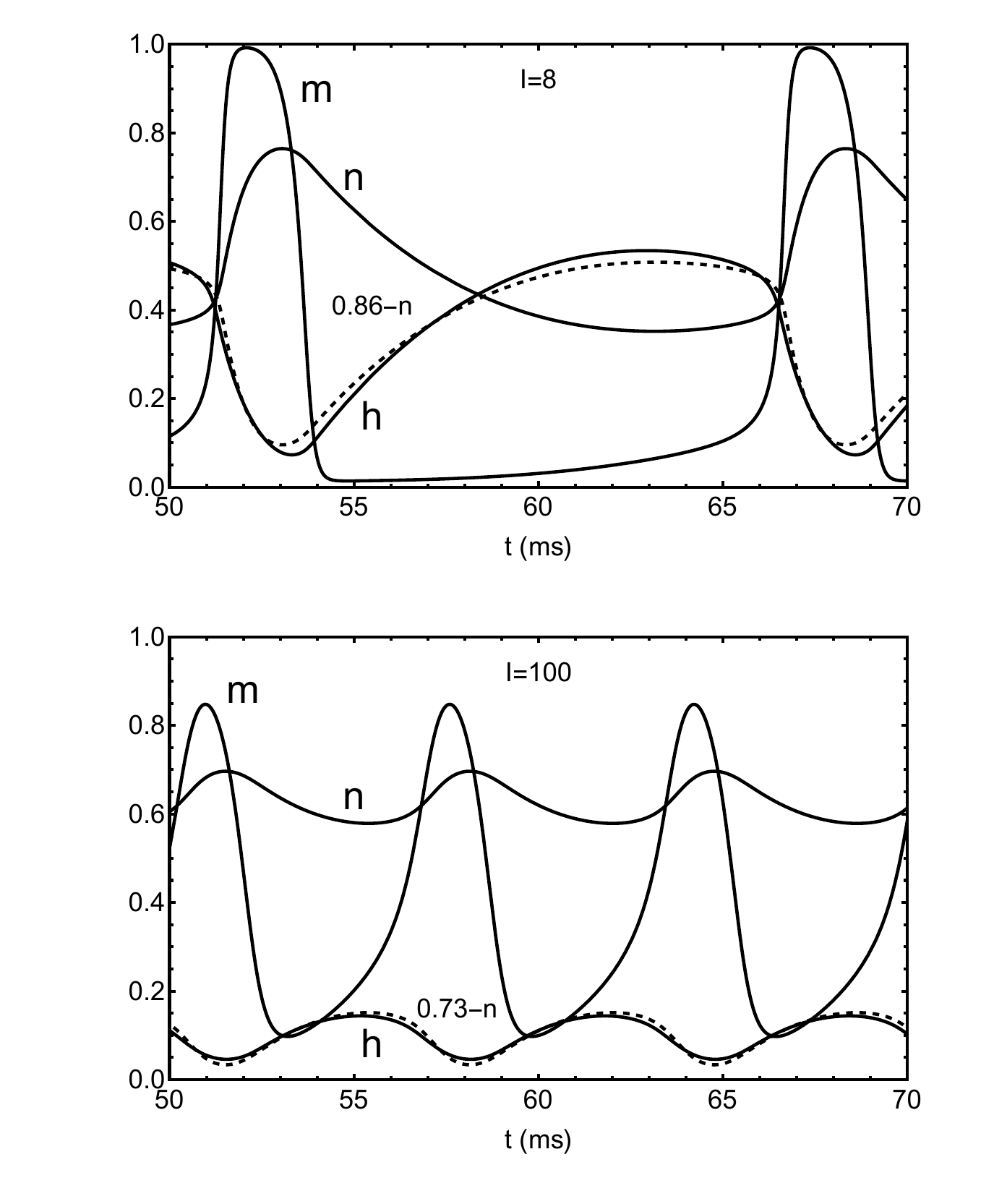}
  \caption{Relationship between $n$ and $h$ for the spiking regimes $I=8$ and $I=100$, obtained with 4D HH equations \eqref{eq-HH}-\eqref{eq-HH2}. The other parameters of the simulations are the same as in Figure \ref{fig:fig1}. For $I=8$, the best fit of $h$ to $h=c_{\hbox{\tiny 3D}}(I)-n$ is for the choice $c_{\hbox{\tiny 3D}}=0.86$ (dashed line). For $I=100$ we obtained   $c_{\hbox{\tiny 3D}}=0.73$ (dashed line).  For other choices of $I$, $c_{\hbox{\tiny 3D}}(I)$ is well described by \eqref{parc}. The gating variables $n$, $m$, and $h$ are dimensionless and take values in the interval $[0,1]$.}
  \label{fig:fig2}
\end{figure}

To analyse the similarities between the solutions of equations \eqref{eq-HH}-\eqref{eq-HH2} and \eqref{eq:3dmodel}, we calculated the values of $c$ that best approximate the function $h=c-n$, obtained with the HH model for several values of $I$ (see Figure \ref{fig:fig2}). A straightforward least squares estimate using numerical solutions of equations \eqref{eq-HH}-\eqref{eq-HH2} and \eqref{eq:3dmodel} shows that the solutions of the 3D reduced HH model \eqref{eq:3dmodel} well approximates the solutions of equations \eqref{eq-HH}-\eqref{eq-HH2} for
\begin{equation}
c=c_{\hbox{\tiny 3D}}(I)= \left\{
\begin{array}{ll}
1.0\ \ &\hbox{for}\ \ I\le 1\\
1.0\, I^{-0.0674}\ \ &\hbox{for}\ \ I>1.
\end{array}\right.
\label{parc}
\end{equation}
showing the dependence of $c_{\hbox{\tiny 3D}}$ on the input current $I$.

Furthermore, we investigated the impact of approximating the gating variables $n$ and $m$ by their steady-state values on the system's behaviour. Approximating $n$ by $n_{\infty}$, where $n_{\infty}(V)$ denotes the steady-state value obtained from $\frac{dn}{dt}=0$,
led to models without Hopf bifurcations and periodic cycles, and all the fixed points are stable (\cite{Bra}). 

The time-dependent gating variable $m(t)$ is closely approximated by
$m_{\infty}$, where $m_{\infty}(V)$ denotes the steady-state value derived from setting 
$\frac{dm}{dt}=0$, best captures the oscillatory dynamics of the 4D  HH model. 
The steady-state solution $m(t)$ is closely approximated in both amplitude and period by $m_{\infty}(V(t))$.
Therefore, we further simplify the 3D HH  model \eqref{eq:3dmodel} to the 2D HH model
\begin{equation}
\begin{array}{rcl}
C_m\frac{dV}{dt} &=& I - g_{\hbox{\tiny K}}n^4(V-V_{\hbox{\tiny K}}) \\[8pt] \displaystyle
&&- g_{\hbox{\tiny Na}}m_{\infty}^3(c_{\hbox{\tiny 2D}}(I)-n)(V-V_{\hbox{\tiny Na}})\\[8pt] \displaystyle
\frac{dn}{dt}  &=& \displaystyle  \alpha_n(V)(1-n)-\beta_n(V)n\\[8pt]  \displaystyle
m_{\infty}(V) &=&\displaystyle  \frac{\alpha_m(V)}{\alpha_m(V)+\beta_m(V)},
\end{array}\label{eq:2dmodel}
\end{equation}
where
\begin{equation}
c_{\hbox{\tiny 2D}}(I)= \left\{
\begin{array}{ll}
1.0\ \ &\hbox{for}\ \ I\le 1\\
1.0\, I^{-0.078}\ \ &\hbox{for}\ \ I>1.
\end{array}\right.
\label{parc2}
\end{equation}

\begin{figure}[!htb]
\centering
\includegraphics[width=0.45\textwidth]{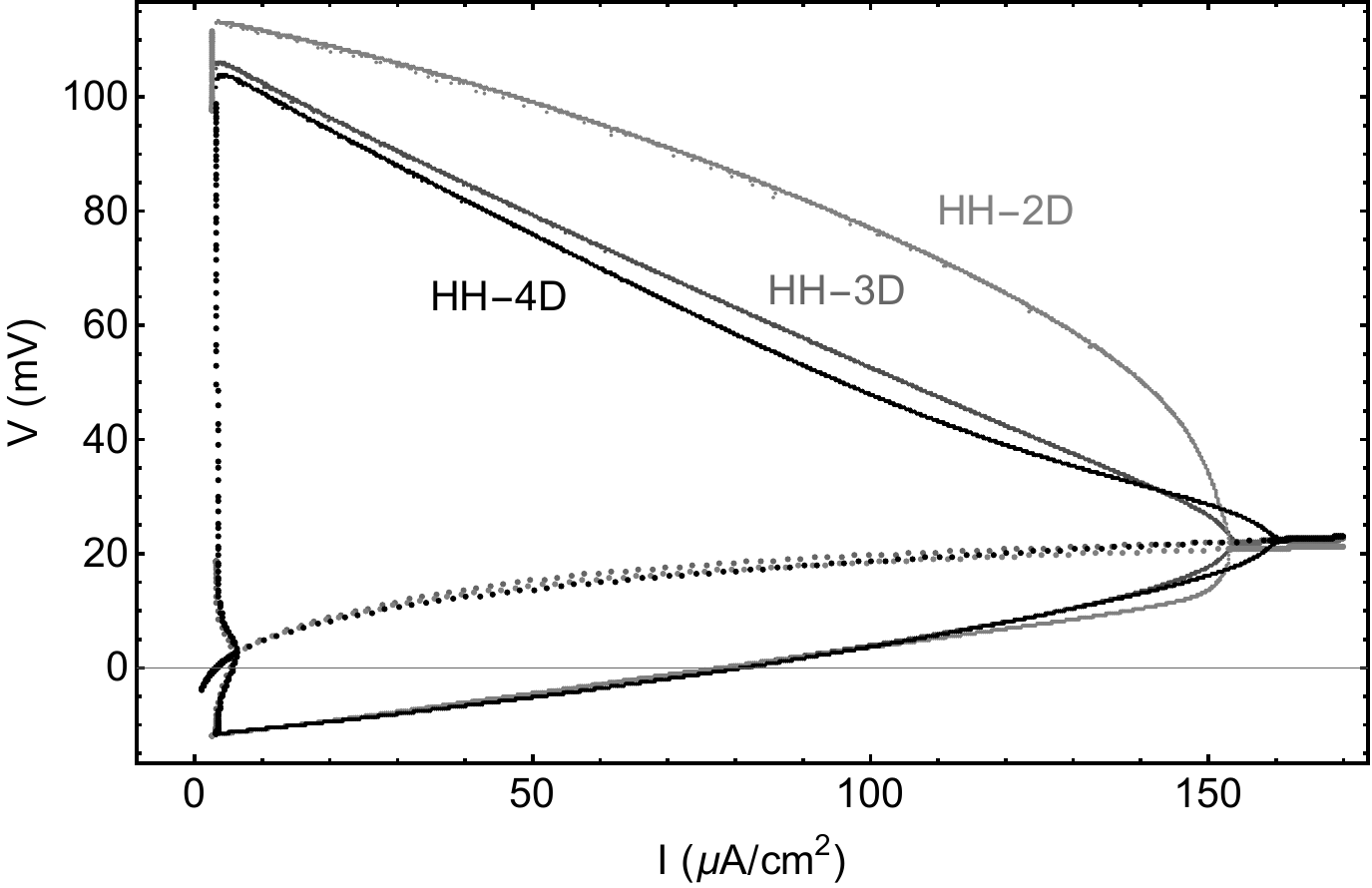}
  \caption{Bifurcation diagrams of the 2D (light grey), 3D (dark grey) and 4D (black) HH models. In the 3D and 2D models \eqref{eq:3dmodel} and \eqref{eq:2dmodel},   the parameter $c=c(I)$  is given by \eqref{parc}  and \eqref{parc2}, respectively.}
  \label{fig:fig3}
\end{figure}

The values of $c_{\hbox{\tiny 3D}}(I)$ and $c_{\hbox{\tiny 2D}}(I)$ were calibrated so that the bifurcation diagrams of the 3D and 2D models closely match those of the 4D model, with similar bifurcation points (Figure \ref{fig:fig3}).

Figure \ref{fig:fig3} depicts the bifurcation diagrams of the three models \eqref{eq-HH}-\eqref{eq-HH2}, \eqref{eq:3dmodel} and \eqref{eq:2dmodel}, with $c_{(\cdot)}(I)$ given by \eqref{parc} and \eqref{parc2}, respectively. 
The similarities between the bifurcation diagrams of the 4D HH model and the reduced models suggest that the behaviour of the Na$^+$ channels can be described by a combination of the gating variables $m$ and $n$. FitzHugh and Rinzel recognised this feature. While FitzHugh considered $c=0.85$ and Rinzel $c=1$, independently of the current density $I$, our analysis shows that $c(I)$ is current-dependent and is better approximated by \eqref{parc} or \eqref{parc2}. Both the 2D and 3D reduced models (equations \eqref{eq:3dmodel} and \eqref{eq:2dmodel}) preserve the main dynamical features of the 4D HH model while reducing its complexity. 
The bifurcation parameters for the 2D and the 3D HH models are: (2D Model) $I_{\hbox{\tiny SNLC}}=2.61$, $I_1=5.31$ and $I_2=153.13$;  (3D Model) $I_{\hbox{\tiny SNLC}}=3.16$, $I_1=5.60$ and $I_2=153.32$. 

We have compared the periods of the action potential spikes in the oscillatory regions $[I_1,I_2-\varepsilon ]$ of the 4D, 3D and 2D HH models, where $\varepsilon=0.02$. For these cases, the periods are well-fitted by the functions
$$\begin{array}{lcl}
 per_{\hbox{\tiny 4D}}(I)&=&32.96\, I^{-0.35}\ \hbox{ms}\\
 per_{\hbox{\tiny 3D}}(I)&=&34.23\, I^{-0.37}\ \hbox{ms} \\
 per_{\hbox{\tiny 2D}}(I)&=&37.33\, I^{-0.49}\ \hbox{ms}.
 \end{array}
$$

The 4D HH model and the 3D and 2D reduced models have a bifurcation structure described by a codimension 2 Bautin bifurcation scenario, \cite{CD1} and \cite{Izh}.

\section{Dynamics of the cable equations}\label{sec3}

\begin{figure}[!ht]
  \centering
  \includegraphics[width=0.5\textwidth]{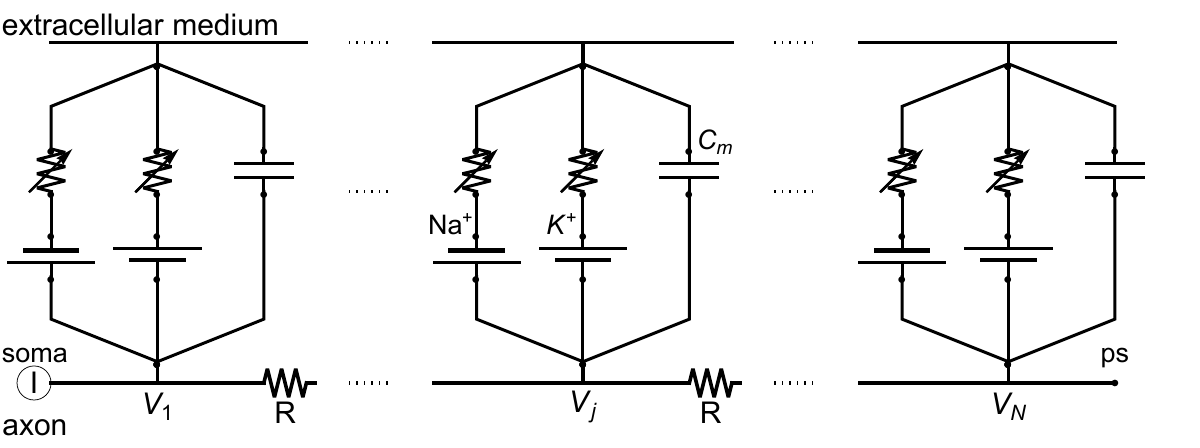}
  \caption{Electric analogue of an axon according to the Hodgkin-Huxley view, excluding leak channels (\cite{HH2}). Each group of active channels communicates with the extracellular space through Na$^+$ and K$^+$ active channels. The signal from the soma is simulated as a current source $I$, and communication between active channels is characterised by the internal resistivity $R$. The presynaptic terminal of the axon is labelled as ``ps''. The membrane potential at each pair of active channels is $V_j$, and the extracellular space potential $V_{\hbox{\tiny out}}$ is assumed constant. The variable resistances of the channels are the inverse of the variable conductances $g_{\hbox{\tiny Na}}m_{\infty}^3(c_{(.)}-n)$ and $g_{\hbox{\tiny K}}n^4$.}
  \label{fig:fig4}
\end{figure}

We represent the axon as a sequence of electrically coupled segments, each containing voltage-gated sodium and potassium channels. Each segment is connected to the neighbouring segments through junctions characterised by an internal resistivity parameter $R$ (\cite{End}). The current density $I$ from the soma of a neuron is transmitted to the first node of its axon, propagating along the sequence of contiguous segments. In Figure \ref{fig:fig4}, we schematically illustrate the electric analogue of an axon for the case where it receives the stimulus from the soma.

Each axonal segment is characterised by a membrane potential  $V_j$, and the axial current between adjacent segments obeys Ohm's law. Under these conditions, and taking as local dynamics equations of the reduced 2D model \eqref{eq:2dmodel}, the equations describing signal propagation along the axon are
\begin{equation}
	\begin{array}{rcl}\displaystyle
	C_m \frac{dV_0}{dt} &=&\displaystyle I + \bigg(\frac{V_1-V_0}{R}\bigg) - I_{ions}(V_0, n_0,m_{\infty}(V_0),c_{\hbox{\tiny 2D}}(I))\\ [8pt]\displaystyle
	C_m \frac{dV_j}{dt} &=&\displaystyle \bigg(\frac{V_{j-1}-V_{j}}{R}\bigg)+ \bigg(\frac{V_{j+1}-V_{j}}{R}\bigg)\\ [8pt]\displaystyle
	&&  -I_{ions}(V_j, n_j,m_{\infty}(V_j),c_{\hbox{\tiny 2D}}(0)),\ \ 0<j<N \\ [8pt]\displaystyle
	C_m \frac{dV_N}{dt} &=&\displaystyle \bigg(\frac{V_{N-1}-V_{N}}{R}\bigg) - I_{ions}(V_N, n_N,m_{\infty}(V_N),c_{\hbox{\tiny 2D}}(0))\\ [8pt]\displaystyle
	\frac{dn_j}{dt} &=&\displaystyle \alpha_n(V_j)(1-n_j)-\beta_n(V_j)n_j ,\ \ 0\le j\le N\\ [8pt]\displaystyle
	m_{\infty}(V) &=&\displaystyle  \frac{\alpha_m(V)}{\alpha_m(V)+\beta_m(V)} ,\ \ 0\le j\le N\\ [8pt]\displaystyle
	I_{ions} &=&\displaystyle g_{\hbox{\tiny K}}n_j^4(V_j-V_{\hbox{\tiny K}}) \\  [8pt]\displaystyle
	&&+ g_{\hbox{\tiny Na}}m_{\infty}(V_j)^3(c_{\hbox{\tiny 2D}}(I\delta_{j0})-n_j)(V_j-V_{\hbox{\tiny Na}}) ,
\end{array}
\label{propagation}
\end{equation}
where $c_{\hbox{\tiny 2D}}(I)$ is defined in \eqref{parc2}, and the $\alpha_{(\cdot)}$ and $\beta_{(\cdot)}$ functions are the usual HH functions defined in \eqref{eq-HH2}, $V_1$ is the potential at the soma-adjacent component of the axon. 
The axon has length $L=N\ell$, where $\ell$ is the length of each voltage-gated element of the axon. 
The soma is at  $x=0$, and the presynaptic terminal is at $x=L$. We use no-flux boundary conditions at $j=0$ and $j=N$. The two coupling terms in the second equation approximate the axial diffusion term $D\frac{\partial^2 V_j}{\partial x^2}=\lim_{\ell\to 0}(V_{j+1}+V_{j-1}-2V_{j})/R$, where $R$ is the internal resistivity of the axon with dimensions $\Omega$cm and $D=\ell^2/R$ is the diffusion coefficient with physical dimensions $S=\Omega^{-1}$. As $\ell\to 0$,  equation \eqref{propagation} converges to a reaction-diffusion equation with zero flux (Neumann) boundary conditions.

Equation \eqref{propagation} has been derived from the 2D HH model with the explicit introduction of the empirical fact that sodium and potassium voltage-gated channels are not dynamically independent, as shown by the relationship between $h$ and $(c_{\hbox{\tiny 2D}}(I)-n)$. As our analysis is primarily focused on cases where the axon excitation occurs at the soma, we have $I=I\delta_{j0}$, where $\delta_{j0}$ is the Kronecker delta function, and outside the soma $I=0$.

For the 4D and 3D HH models, we use equations \eqref{eq-HH}-\eqref{eq-HH2} and \eqref{eq:3dmodel} with the nearest-neighbour coupling as in \eqref{propagation}.
In the following subsections, we compare the solutions of three models (4D, 3D and 2D) and derive their propagation properties as a function of the axon-internal resistivity $R$ and membrane capacitance $C_m$.

\subsection{Signal transmission along the axon -- 4D model}

\begin{figure}
\centering
\includegraphics[width=0.35\textwidth]{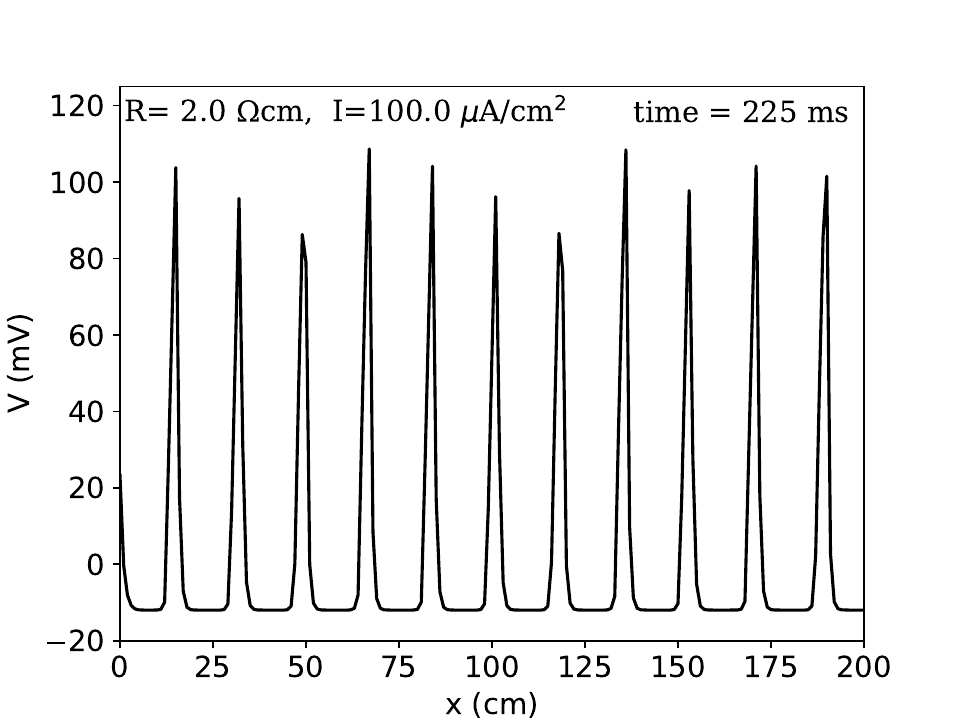}
\includegraphics[width=0.35\textwidth]{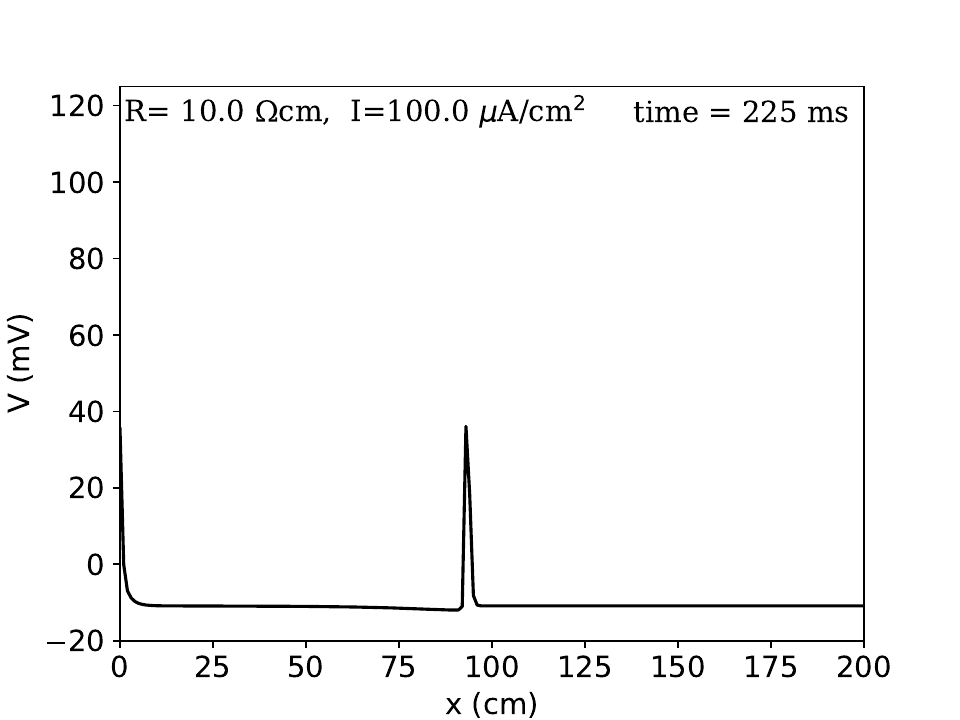}
 \caption{Transmembrane potential along an axon at time $t=225$~ms, for a constant stimulus applied at the soma $I=100$~$\mu$A/cm$^2$, described by the 4D HH model \eqref{eq-HH}-\eqref{eq-HH2}, with $g_L =0$. For $R=2.0$~$\Omega$cm, the action potential propagates along the axon as an approximately periodic time function.
 For $R=10.0$~$\Omega$cm, a solitary spiky signal forms and propagates along the axon; after colliding with the presynaptic region, a constant steady state is established, and no further axonal activity is observed.}
 \label{fig:fig5}
\end{figure}

The signals generated by the stimulus current entering from the soma travel unidirectionally along the axon. For later comparison and reference, we use the 4D HH  model \eqref{eq-HH}-\eqref{eq-HH2}, with $g_L =0$, internal resistivity $R$ and boundary conditions calculated as in model equations \eqref{propagation}.  Figure~\ref{fig:fig5} depicts two action-potential signal solutions along the axon at time $t=225$~ms, for a soma ($x=0$) constant stimulus $I=100$~$\mu$A/cm$^2$, and several values of the internal resistivity $R$.  We have considered an axon with $N=200$ axon segments, each with length $\ell =1$~cm. The equations \eqref{propagation} were integrated with the Euler method with $dt=0.001$~ms. The initial conditions at the steady state  $I=0$ are $$(V^*,n^*,m^*,h^*)=(-10.8781, 0.1710, 0.0138, 0.8796).$$

\begin{figure}
\centering
\includegraphics[width=0.38\textwidth]{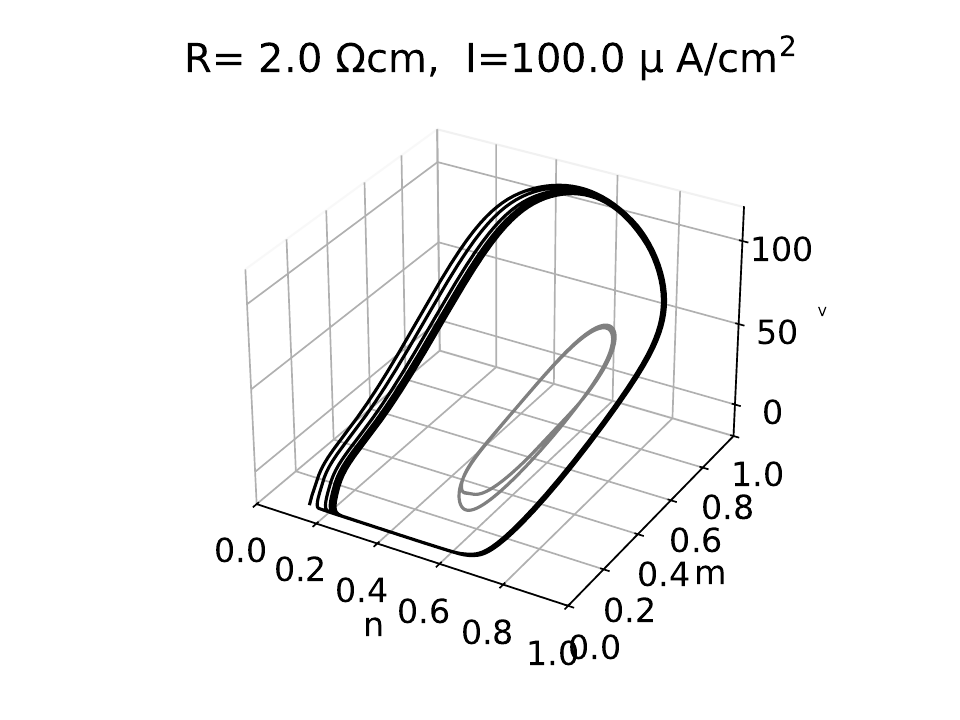}
\caption{Projection of the phase space orbits of the 4D HH model \eqref{eq-HH}-\eqref{eq-HH2}, with $g_L =0$, at the soma  ($j=0$, in grey) and the middle of the axon ($j=100$, in black), during the time interval $[500,2000]$~ms.}
 \label{fig:fig6}
\end{figure}

Within $R\in [0.01,20]$, the response exhibits several dynamical regimes.
For $R\lnapprox 0.01$~$\Omega$cm or $R\gnapprox 18$~$\Omega$cm, the axon remains at a steady state, and the action potential does not propagate. For $0.01\lnapprox R\lnapprox0.02$ or $4.0\lnapprox R \lnapprox 17.5$, a solitary action potential spike is produced and propagates along the axon before the axon settles into a steady state. For $0.02\lnapprox R\lnapprox 3.9$, periodic trains of action potentials form and propagate along the axon.

Figure~\ref{fig:fig6} displays two three-dimensional phase-space projections of the asymptotic orbits from solutions to the 4D HH model, with $R=2.0$~$\Omega$cm and $I=100$$\mu$A/cm$^2$, at the soma and mid-axon.

\begin{figure}
\centering
\includegraphics[width=0.3\textwidth]{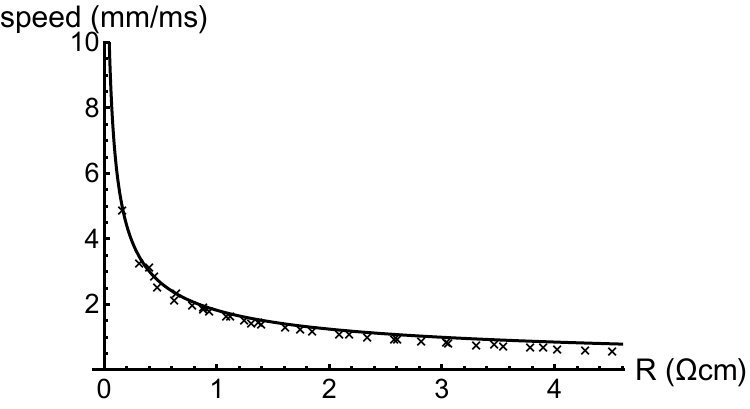}
\caption{Propagation speed of the first action potential spike as a function of the internal resistivity $R$, for $I=20,100, 140$~$\mu$A/cm$^2$ and $C_m = 1$~$\mu F/$cm$^2$, calculated from the 4D HH model (1)-(2). The action-potential propagation speed is well approximated by equation \eqref{speed2-4D}. The fitted functions \eqref{speed-4D} are practically indistinguishable from \eqref{speed2-4D}.}
 \label{fig:fig7}
\end{figure}

If an action-potential spike is produced, the first spike propagates along the axon at approximately constant speed (\cite{CD2}). The propagation speed of the initial action potential's spike was numerically computed for different somatic stimulus values $I$ as a function of the internal resistivity $R$. While the propagation speed depends on the internal resistivity of the axon (see Figure~\ref{fig:fig7}), it is only minimally influenced by the stimulus $I$ at the soma. The speed of the first spike can be precisely characterised by the following functions
\begin{equation}
\begin{array}{lcl}
v_{\hbox{\tiny 4D}}(R,I=20)=1.83\ R^{-0.63}, \ R\in [0.4,4.5]\\
v_{\hbox{\tiny 4D}}(R,I=100)=1.81\ R^{-0.54}, \ R\in [0.02,3.9]\\
v_{\hbox{\tiny 4D}}(R,I=140)=1.82\ R^{-0.54}, \ R\in [0.01,1.4],
\end{array}
\label{speed-4D}
\end{equation}
with units of mm/ms.

This suggests that the propagation speed of the action potential along the axon is approximately described by the function
\begin{equation}
v_{\hbox{{\tiny 4D}}}(R)=a \frac{1}{R^{b}} \ \ \hbox{mm/ms},
\label{speed2-4D}
\end{equation}
where $a=1.82$ and $b=0.55$. 

For $I=20$~$\mu$A/cm$^2$, oscillatory responses along the axon are in the internal resistivity interval $0.4\lnapprox R\lnapprox 10.0$. For $I=140$~$\mu$A/cm$^2$, we obtained oscillatory responses for   $R\in [0.01, 1.4]$.

\begin{figure}
\centering
\includegraphics[width=0.40\textwidth]{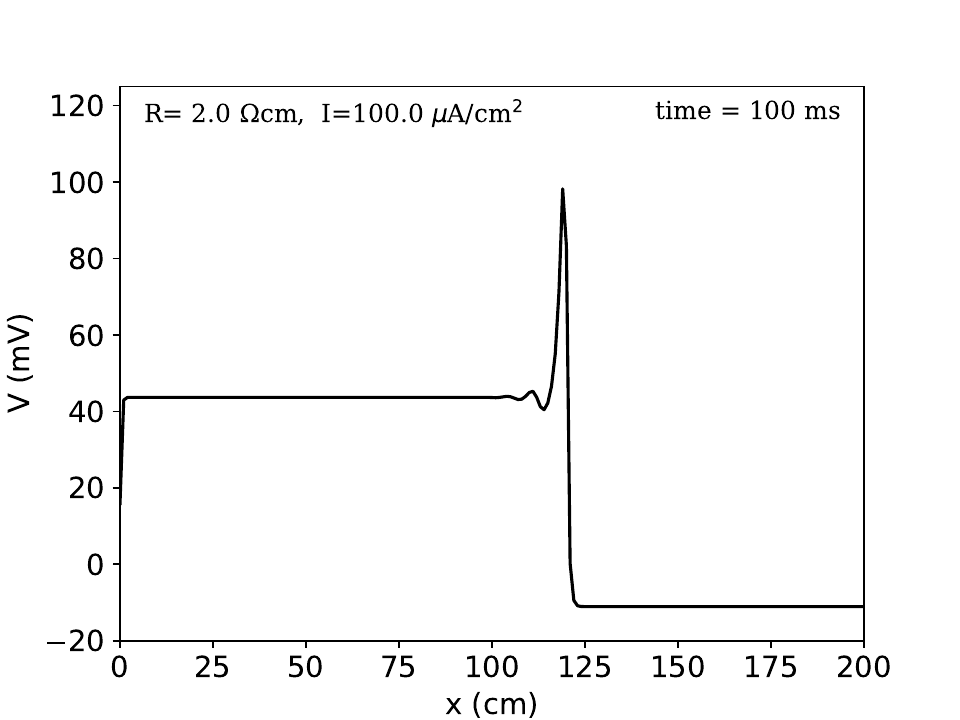}
\caption{Action potential solitary spike solution for the 3D HH model, computed using the spatially extended model equations \eqref{propagation}  adapted to the reduced 3D HH model and the parameterisation \eqref{parc}.}
 \label{fig:fig8}
\end{figure}

\begin{figure*}
\centering
\includegraphics[width=0.8\textwidth]{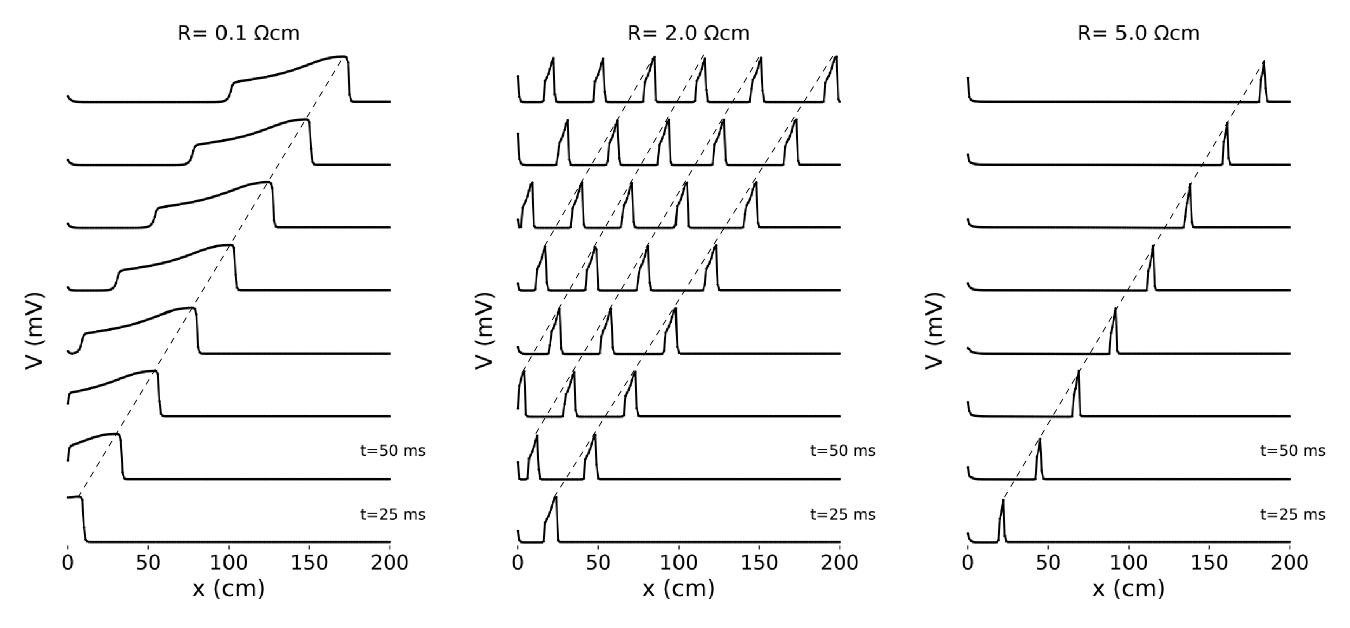}
 \caption{Propagation of the transmembrane potential along the axon with a constant stimulus at the soma with $I=100$~$\mu$A/cm$^2$ and $C_m = 1$~$\mu F/$cm$^2$, for several values of the internal resistivity $R$, computed using the spatially extended model equations \eqref{propagation} relative to the  2D HH model. The vertical axis range is $[-20,125]$~mV. The axon has $N=200$ segments and length $L=2$~m. The dashed lines are the characteristic curves of the action potential spikes, showing that the propagation speed of each spike is constant. For $R=5.0$~$\Omega$cm, we have a solitary spike solution that annihilates at the presynaptic terminal of the axon, and for $t>200$~ms, the axon stays in a steady state.}
 \label{fig:fig9}
\end{figure*}

\subsection{Signal transmission along the axon -- 3D model}

The same approach followed for the 4D HH model with $g_L =0$ can be tested for the 3D model with $h=c_{\hbox{\tiny 3D}}(I)-n$, and $c_{\hbox{\tiny 3D}}(I)$ as defined in \eqref{parc}. The bifurcation diagrams for the local 4D and 3D systems are very similar. However, numerical results with the discrete setting \eqref{propagation} give solitary spike solutions with a single spike and the establishment of a constant steady state along the axon for $0.1\lnapprox R\lnapprox 7.0$ (Figure~\ref{fig:fig8}). In this interval, the propagation speed of the solitary action-potential spike is well described by the function $v_{\hbox{\tiny 3D}}=1.81/R^{0.52}$~mm/ms, which is close to the estimate for the 4D model \eqref{speed2-4D}.

\subsection{Signal transmission along the axon -- 2D model}

We now consider the spatially extended 2D HH model  \eqref{propagation}, where $c_{\hbox{\tiny 2D}}(I)$ is defined in \eqref{parc2}, and the $\alpha_{(\cdot)}$ and $\beta_{(\cdot)}$ are the usual HH functions as defined in \eqref{eq-HH2}.  

We consider  the same parameter values of the 4D model but with initial conditions $(V^*,n^*)=(-10.9506, 0.1702)$. In the 2D HH reduced model, for $0.01\lnapprox R\lnapprox 4.6$, action-potential trains propagate along the axon and are periodic in time for any longitudinal position along the axon. For $R< 0.05$ and $R> 4.6$, one spike is produced, and, as time passes, the spatial solution along the axon converges to a steady state (Figure~\ref{fig:fig9}). 

\begin{figure}
\centering
\includegraphics[width=0.23\textwidth]{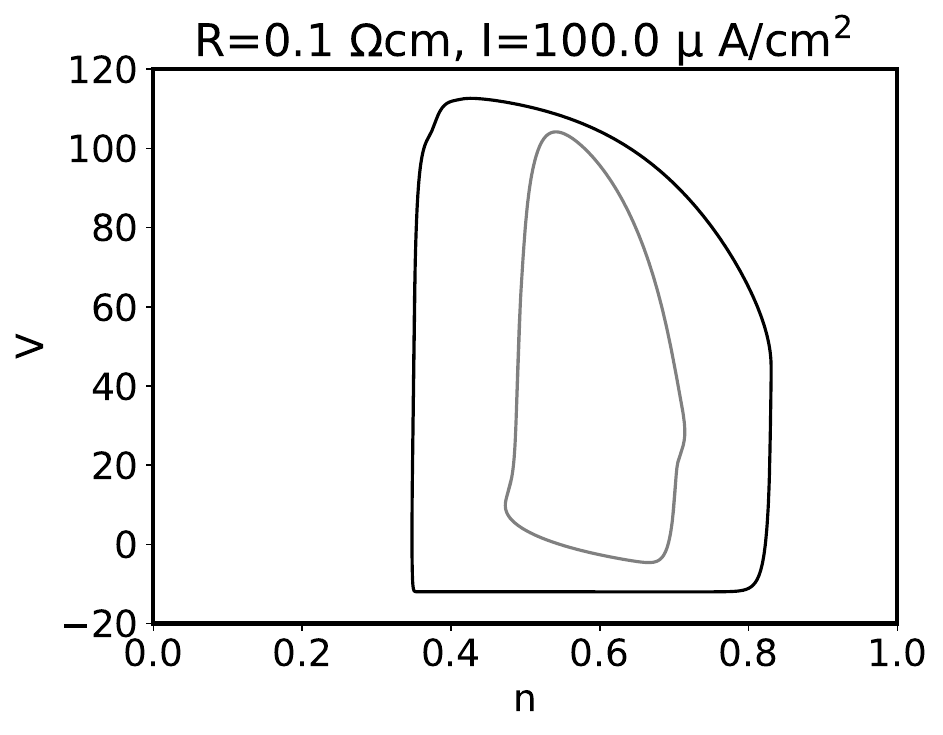}
\includegraphics[width=0.23\textwidth]{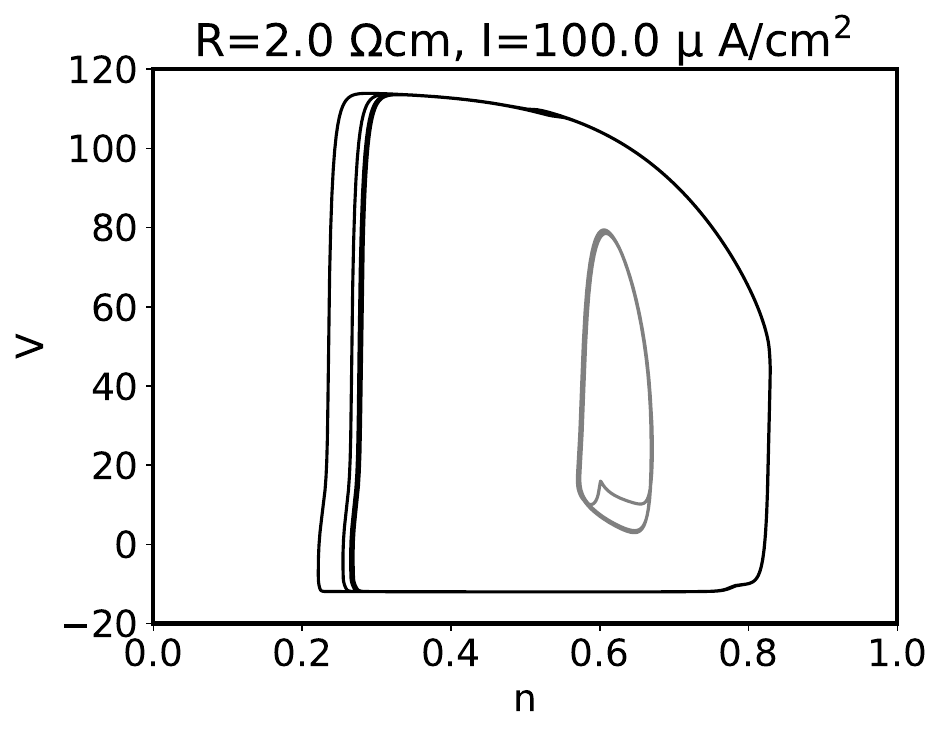}
\caption{Phase space orbits at the soma  ($j=0$, in grey) and at the middle of the axon ($j=100$, in black), obtained during the time interval $[500,2000]$~ms, with the 2D HH model \eqref{propagation}.}
 \label{fig:fig10}
\end{figure}

In Figure~\ref{fig:fig10}, we depict the phase-space orbits at the soma and the axon's middle for two internal resistivity values $R$, showing the difference in dynamics between these axon regions.

The action-potential spikes propagate at approximately constant speed.  As in the 4D model, the propagation speed of the first action-potential spikes is well approximated by the function $v_{\hbox{\tiny 2D}}(R,C_m=1)=4.43 /R^{0.66}$, for $I=100$ and $R\in[0.01,4.6]$. 

The width of the spiky action-potential signals depends on the internal resistivity (Figure~\ref{fig:fig9}). To quantify this effect, we calculate the width at half height, $w$, of the first action potential spike. For $I=100$ and $R\in[0.01,4.6]$, we obtain $w_{\hbox{\tiny 2D}}(R,C_m=1)=9.0 /{R^{0.76}}$~mm.

\begin{figure}[!htb]
  \centering
  \includegraphics[width=0.3\textwidth]{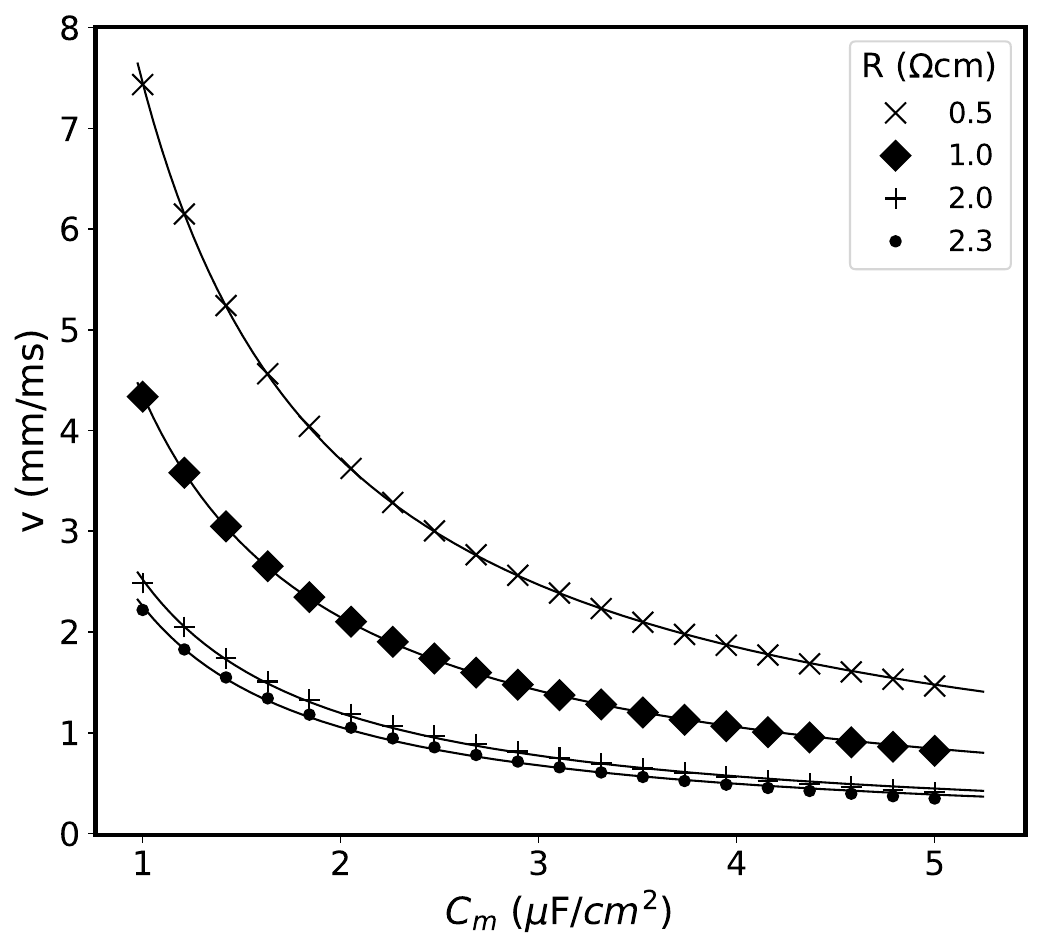}
  \caption{Relation between the propagation velocity $v$ and the membrane capacitance $C_m$ for several internal resistivity values, calculated with the 2D HH model.  The black lines show the fit functions  $v = a_1(R)/C_m^{b_1(R)}$, where the parameter values for $a_1(R)$ and $b_1(R)$ are shown in Table \ref{tab:v(cm)}.}
  \label{fig:fig11}
\end{figure}

To examine the influence of membrane capacitance on the propagation velocity of action potentials, we conducted additional simulations with various values of $C_m$ ranging from $1$ to $5$~$\mu$F/cm$^2$, while keeping some of the previously used internal resistivity values.
We employed a fit function of $v = a_1/C_m^{b_1}$  with an expected value of $b_1 = 1$, as suggested by \cite{KS}. 
The specific values of $ a_1(R)$ and $b_1(R)$ obtained for different internal resistivities are listed in Table~\ref{tab:v(cm)}, and the resulting fit functions, as well as the fitted points, are depicted in Figure \ref{fig:fig11}.

\begin{table} 
\caption{Results of fitting the data points $v(C_m)$ to the function  $v = a_1(R)/C_m^{b_1(R)}$, for various internal resistivities}\label{tab:v(cm)}
  \begin{tabular}{lcccccc}
    \hline
    R $(\Omega cm$)  & 0.5 & 1.0 & 2.0 .& 2.3   \\
  \hline
  $ a_1(R)$  & 7.45 & 4.36 & 2.52 & 2.26   \\
    $b_1(R)$  & 1.01 & 1.02 & 1.08  & 1.10  \\
  \hline
  \end{tabular}
\end{table}

From Table \ref{tab:v(cm)}, the fitted values of $b_1$ show only a weak dependence on $R$ and are close to 1.
The parameter $a_1$ is also well described by a power law in $R$, giving $a_1=4.34/R^{0.78}$. 
Merging  all these fits, it follows that the action potential velocities are well approximated by
$$
v_{\hbox{\tiny 2D}}(R, C_m)=a \frac{1}{C_m R^{b}}
$$
where $a$ and $b<1$ are positive constants.

\subsection{Direction of propagation of action potential signals}

We now examine how the axon responds to a current signal, either from a patch clamp experiment injected at a specific position $0<x_e<L$ of the axon.

\begin{figure}
\centering
\includegraphics[width=0.49\textwidth]{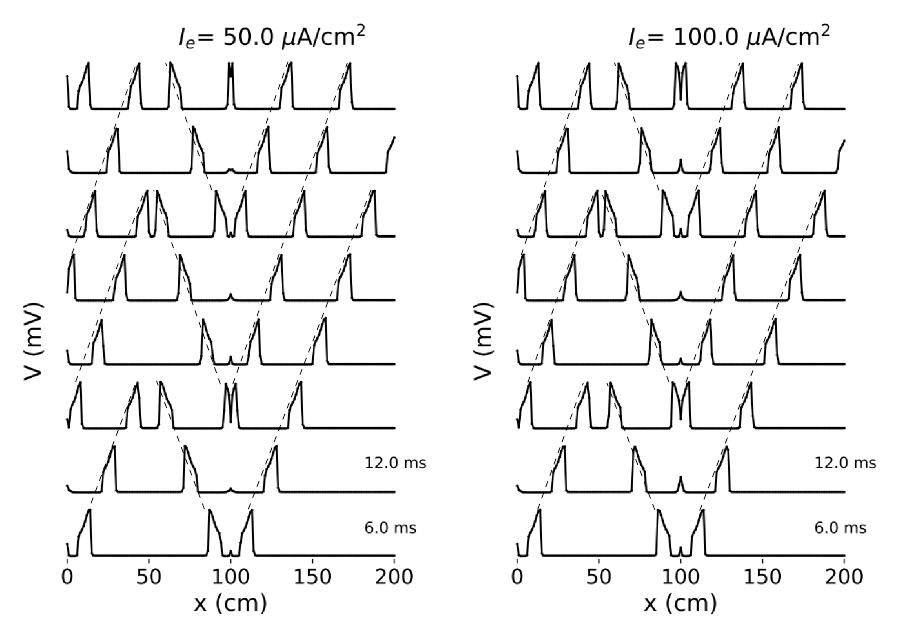}
 \caption{
 Propagation of the transmembrane potential along the axon with a constant stimulus at the soma ($I=100$~$\mu$A/cm$^2$) and a second constant stimulus at $x_e=100$ ($I_e=50$~$\mu$A/cm$^2$ and $I_e=100$~$\mu$A/cm$^2$), calculated with the 2D HH model. Other simulation parameters are $C_m = 1$~$\mu F/$cm$^2$ and $R=2$~$\Omega$cm. The axon has $N=200$ segments and length $L=2$~m, and the vertical axis range is $[-20,125]$~mV.  
The dashed lines are the characteristic curves of the action potential spikes with a slope proportional to the speed of the spikes. The signal injected at $x_e=100$ annihilates the signal arriving from the soma. The signal that reaches the presynaptic region of the axon is generated in the middle of the axon.}
 \label{fig:fig12}
\end{figure}

In Figure~\ref{fig:fig12}, we depict the temporal evolution of the axon's membrane potential in response to a constant stimulus $I=100$~$\mu$A/cm$^2$ originating from the soma, along with localised stimuli at $x_e=100$ with $I_e = 50$ (left) and $I_e = 100$ (right). We considered that $R=2$~$\Omega$cm and $C_m = 1$~$\mu F/$cm$^2$. In these simulations, the equation for $V_j$, when $j=e$, includes the terms $I_e\delta_{je}$ and $c(I_{j=e})=c(I_e\delta_{je})$.

In these simulations, two action potential train spikes are produced at $x_e=100$, propagating in opposite directions along the axon. One of the spikes annihilates the membrane potential response generated at the soma, while the other action potential spike produced at the branching point propagates towards the presynaptic region of the axon. Similar effects were also observed in the 4D HH model in extended domains (\cite{CD2}).

\section{Conclusions}\label{sec4}
\label{sec:concl}

We simplified the 4D HH model into new 3D and 2D models while maintaining its dynamical and electrophysiological characteristics, as well as the bifurcation scenario. This method contrasts with the heuristic reductions suggested by \cite{FH} and \cite{AK}. A key step in this reduction replaces the sodium gating variable $h$ in the 4D HH model with $c(I)-n$, where $I$ represents the electric stimulus intensity from the soma. The literature also reports a similar empirical relationship between the gating variables $h$ and $n$, assuming that $c$ remains a constant independent of the electric stimulus $I$ at the soma, \cite{FH}, \cite{Rin2},  \cite{KS} and \cite{Wan}. 

Our analysis shows that action potential propagation happens within a particular internal resistivity range, $R\in [R_{m0}, R_{M0}]$. Beyond these limits, the axon remains in a steady state regardless of the soma's (current) signal intensity. Within the internal resistivity interval $[R_{m0},R_{M0}]$, there is a narrower interval $[R_{m1},R_{M1}]\subset [R_{m0},R_{M0}]$ where periodic or nearly periodic action potentials are generated at each spatial position along the axon. For $R\in [R_{m0},R_{M0}]-[R_{m1},R_{M1}]$, the sustained current from the soma causes a single action potential spike to travel along the axon, terminating at the presynaptic terminal, after which the axon returns to its resting state.

For the $4D$, $3D$ and $2D$ HH   models, the propagation speed of the action-potential response is well-described by the function
\[
v(R, C_m)\simeq a  \frac{1}{C_m R^{b}}
\]
where $a$ and $b <1$ are positive constants independent of the soma stimulus intensity.  For the three models, we have obtained $b=0.55$ (4D), $b=0.52$ (3D) and $b=0.66$ (2D), and  the width of the action potential spikes also depends on 
the internal resistivity with $w(R, C_m=1)=a_2 /R^{b_2}$, where $a_2$ and $b_2$ are positive constants.

The simplicity of the 2D model may enable more computationally efficient simulations of neuronal circuits and networks.

\section*{Data availability statement}
No data is associated with the manuscript.


\end{document}